\documentclass{iaus}
\usepackage{graphicx}

  \checkfont{eurm10}
  \iffontfound
    \IfFileExists{upmath.sty}
      {\typeout{^^JFound AMS Euler Roman fonts on the system,
                   using the 'upmath' package.^^J}%
       \usepackage{upmath}}
      {\typeout{^^JFound AMS Euler Roman fonts on the system, but you
                   dont seem to have the}%
       \typeout{'upmath' package installed. iaus.cls can take advantage
                 of these fonts,^^Jif you use 'upmath' package.^^J}%
      }
  \else
  \fi

  \checkfont{msam10}
  \iffontfound
    \IfFileExists{amssymb.sty}
      {\typeout{^^JFound AMS Symbol fonts on the system, using the
                'amssymb' package.^^J}%
       \usepackage{amssymb}%

      }{}
  \fi

  \IfFileExists{amsbsy.sty}
    {\typeout{^^JFound the 'amsbsy' package on the system, using it.^^J}%
     \usepackage{amsbsy}}
    {\providecommand\boldsymbol[1]{\mbox{\boldmath $##1$}}}

\newsavebox{\astrutbox}
\sbox{\astrutbox}{\rule[-5pt]{0pt}{20pt}}

\newcommand\etal{\mbox{\textit{et al.}}}

\title[The Interplay among Black Holes, Stars and ISM in Galactic
       Nuclei]{Estimating the Ages of Bars: Implications for the Bar-AGN-Star Formation Connection}

\author[D. A. Gadotti and R. E. de Souza]
{Dimitri Alexei Gadotti \and Ronaldo Eust\'aquio de Souza}

\affiliation{Departamento de Astronomia, Universidade de S\~ao Paulo, Rua do Mat\~ao, 1226,
05508-900, S\~ao Paulo-SP, Brasil email: dimitri@astro.iag.usp.br,ronaldo@astro.iag.usp.br}

\pubyear{2004}
\volume{222}
\pagerange{1--8}
\date{?? and in revised form ??}
\jname{The Interplay among Black Holes, Stars and ISM \\in Galactic Nuclei}
\editors{Th. Storchi Bergmann, L.C. Ho \& H.R. Schmitt, eds.}
\begin{document}

\maketitle

\begin{abstract}
In an effort to elevate to higher grounds our understanding on the impact of the formation and evolution
of bars in the formation and evolution of galaxies, we have developed a diagnostic tool to distinguish
between recently formed and evolved bars. Our method was applied in the study of a sample of 14
galaxies and revealed that, apparently, AGN activity tends to appear in galaxies which have young
bars rather than evolved bars. This suggests that the time scale for the fueling of AGN by bars is short,
and may help to explain, for instance, why there is not a clear correlation between the presence of
bars and AGN in galaxies.
\end{abstract}

\firstsection 

\section{Motivation}

It has long been known that once a bar develops in a galaxy it induces several evolutionary
processes which severely modify many aspects of its host. Changes include
the stellar and gas dynamics, by the onset of new relevant orbits, resonances and torques
\cite[(see, e.g., Sellwood \& Wilkinson 1993; Combes 2001)]{Sellwood93,Combes01};
the evolution of the stellar populations, by the attenuation of chemical composition gradients
\cite[(Martin \& Roy 1994; Zaritsky, Kennicutt \& Huchra 1994)]{Martin94,Zaritsky94} and
the triggering of nuclear starbursts \cite[(e.g., Gadotti \& dos Anjos 2001)]{Gadotti01};
and the global galactic structure, with the formation of rings and lenses, and the building of bulges
\cite[(see Athanassoula 1983; Buta \& Combes 1996; Carollo,
this volume)]{Athanassoula83,Buta96}. Bars also are very likely related to the problem of fueling
AGN (see, e.g., Combes, this volume, and references therein)
transferring outwards the angular momentum of gas and
stars in the disk. However, these processes may have different time scales and, what is particularly
relevant for the subject of this meeting, this may explain, at least partially, the absence of a clear
correlation between the presence of bars and AGN in galaxies. For instance, a barred galaxy may not
have an AGN simply because its bar did not yet have the time to funnel the disk material to the
center. Inversely, if the time scale for AGN fueling by bars is short, then we may be observing
barred galaxies which had already their AGN phases. The availability of AGN fuel, the presence
of inner Lindblad resonances and nuclear spirals, the possibility
of destruction and rejuvenation of bars, and the recurrence of AGN phases, add complexity to
the theme. Indeed, it is really naive to expect a clear correlation between bars and AGN given
all the details involved.

While the time scale issue is recognized now as a very important point to better understand
the observations (which means that it is also a key ingredient in theoretical modeling), very
little has been done on this regard, considering bars. In spite of its relevance to many aspects
of the formation and evolution of galaxies, we still lack the means to evaluate the age of a bar, i.e.,
for how long its presence is altering the evolutionary course of its host. Thus, we have developed a
diagnostic tool to distinguish between recently formed and evolved bars, and observationally estimate
the age of a bar in a galaxy. This allows us, for instance, to directly tackle the issue of the time scale
in the AGN fueling by bars.

\section{Method}

Our method to estimate the age of a bar is based on the fact that, supposedly, bars form in disks
\cite[(but see Gadotti \& de Souza 2003)]{Gadotti03b}, which means that they are initially vertically thin, and
then their vertical velocity dispersion, $\sigma_z$, must be initially very small, i.e. a few tens of Km/s. Moreover,
as bars evolve, vertical resonances, and/or the hose instability, give them an important vertical
structure, which may be recognized by higher values for $\sigma_z$, after about 1 Gyr or so
\cite[(see Friedli 1999 and references therein)]{Friedli99}.

Thus, we have taken high S/N spectra (around the MgI 5175 \AA\ feature)
along the bar axes of 14 face-on barred galaxies (with
morphological types ranging from S0 to Sb), which allowed us
to determine $\sigma_z$ after fitting the line of sight stellar velocity distribution, using a technique
similar to the one developed by \cite{vanderMarel93}. Since it is based on Gauss-Hermite series,
allowing for non-gaussian line profiles, this technique provides reliable estimates for the kinematical
parameters. To minimize even further errors coming from template mismatch, we have used template spectra
from many stars of different spectral types. These data were obtained at the 1.52 m ESO telescope at
La Silla, Chile, and at the 2.29 m Steward Observatory telescope at Kitt Peak, Arizona. Details may
be found in \cite{Gadotti03a}.

\begin{figure}
\centering
\includegraphics[width=13cm]{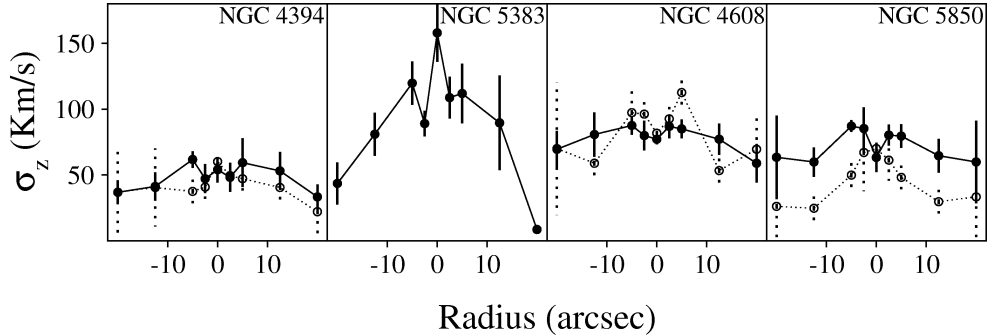}
\caption{Measured values for $\sigma_z$ along the bar major (filled circles, solid lines and error bars)
and minor (open circles, dotted lines and error bars) axes of 4 galaxies in our sample. The
two panels at left are examples of recently formed bars while the ones at right are examples of
evolved bars. In NGC 5383 (for which we haven't taken spectra along the bar minor axis),
the high values of $\sigma_z$ in the inner region correspond to its
bulge, while the drops at each side of the center are caused by its inner spiral arms. In NGC 5850,
an inner bar identified in the {\sc budda} \cite[(de Souza, Gadotti \& dos Anjos 2004)]{deSouza04}
residual images is the likely reason for the noticeable drop in the center.}
\end{figure}

Figure 1 shows some examples on how we can then separate young and evolved bars. In principle, if
the bulge of a galaxy has a high velocity dispersion (which is generally the case), a young bar will have
a $\sigma_z$ radial profile which steeply falls from the center along its major axis.
This is the case of NGC 5383. Note further that the bar of this galaxy has $\sigma_z<50$ Km/s.
On the other hand, evolved bars have a flatter profile, like NGC 4608, whose bar has $\sigma_z>50$ Km/s.
There may be, however, more ambiguous cases. NGC 4394 has a flatter profile, but its bar is
certainly young, as can be concluded from the low $\sigma_z$ values. Its bulge then has
a low pressure support. Another interesting example
is NGC 5850. The spectra obtained along its bar minor axis reach regions where the bar and the bulge
contribution to the light is small, and hence we are measuring the pure disk velocity dispersion. As it is
around a factor of 2 or more lower than along the bar major axis this is certainly an evolved bar, which
is already expected from the flat $\sigma_z$ radial profile. Note that
in e.g. NGC 4608 the minor axis spectra do not escape the bulge and bar influence.

\section{Results and Discussion}

From the 14 galaxies of our sample we found 8 evolved bars and 5 young ones (1 case is not
conclusive). Only 2/8 (25\%) galaxies with evolved bars have AGN, while 3/5 (60\%) galaxies with young
bars have AGN. Then we have to conclude that AGN appear more often in galaxies with
young rather than evolved bars, which in turn suggests that the time scale for AGN feeding by bars
is short. This means that a clearer correlation between the presence of bars and AGN in galaxies
may be observed when using a sample of barred galaxies whose bars are recently formed. This
result is obviously based on a small number statistics and should be confirmed by further investigation.
Note that to determine if a galaxy has a young or evolved bar is a task which is expensive in terms
of telescope time.

We now may ask how young are recently formed bars or what is the age difference between young
and evolved bars. A more difficult task, which needs deep theoretical analysis, is to obtain the age
of the bar from the kinematical measurements. To investigate these questions, we have performed
several $N$-body simulations in which bars form and evolve for 2 Gyr. These experiments consider only
gravity physics and stars, and show that boxy/peanut bulges appear in time scales of less
than about 1 Gyr, and even after 2 Gyr the values for $\sigma_z$ in the bar are still lower than
about 50 Km/s. These results lead us to conclude that, to explain the values we measured for
$\sigma_z$, another mechanism, rather than the ones which originate boxy/peanut bulges,
namely vertical resonances and/or the hose instability, is playing a relevant role to give bars
an expressive vertical structure. We suggest that the
\cite[Spitzer \& Schwarzschild (1951,1953)]{Spitzer51,Spitzer53} mechanism may
be the answer if we assume that giant molecular clouds are twice as much concentrated along
the bar as in the disk. We have analytically estimated that in this case bars reach the observed
$\sigma_z$ in time scales of $5 - 10$ Gyr. These questions were also tackled by multiband (BVRIKs)
imaging, which allowed us to measure the colors of the bars. Interestingly, the mean difference in the
colors of young and evolved bars suggests a difference in age of about the same time scale.
These results will appear in a forthcoming paper, but see \cite{Gadotti03a}.

\begin{figure}
\centering
\includegraphics[width=6.55cm]{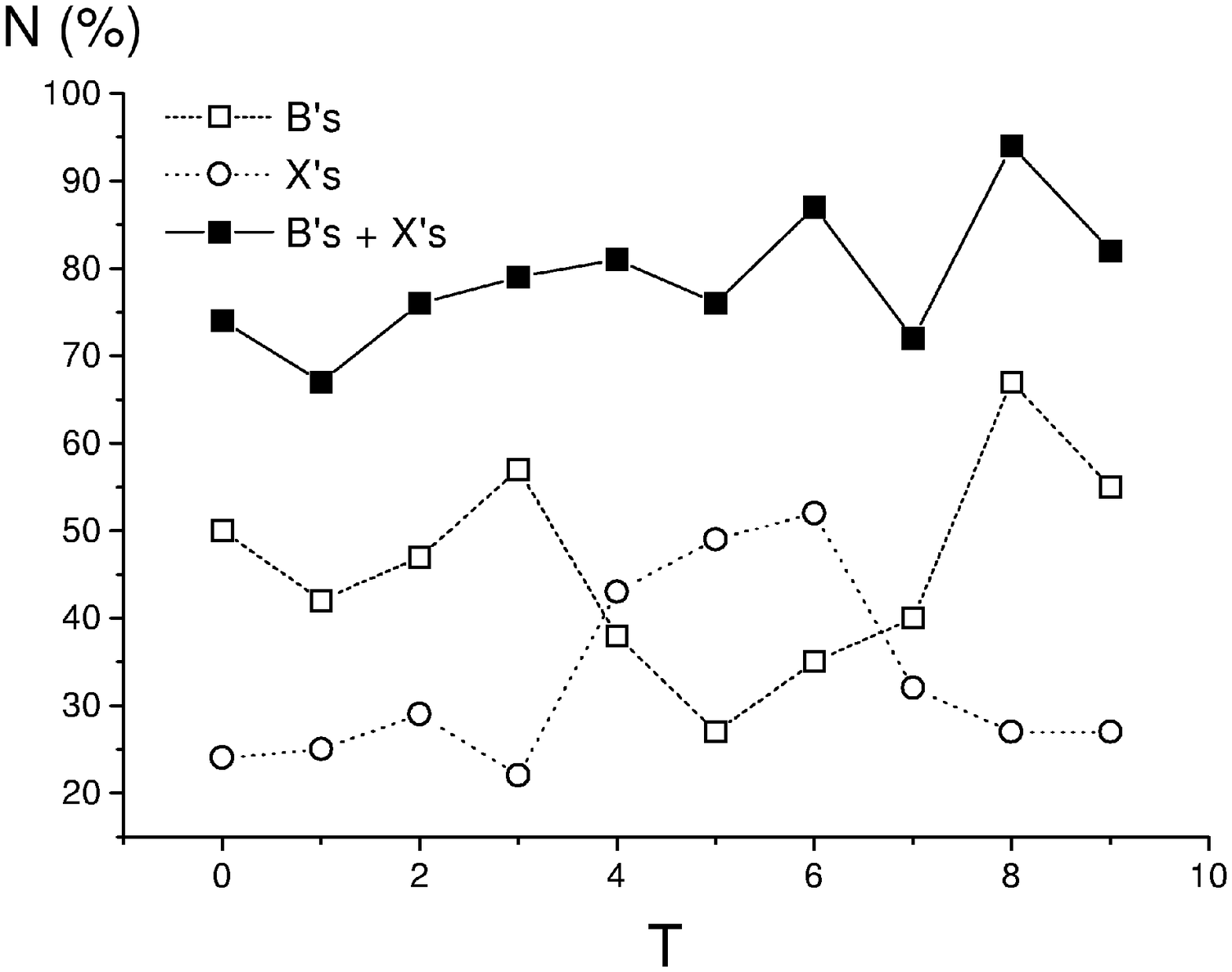}\includegraphics[width=6.45cm]{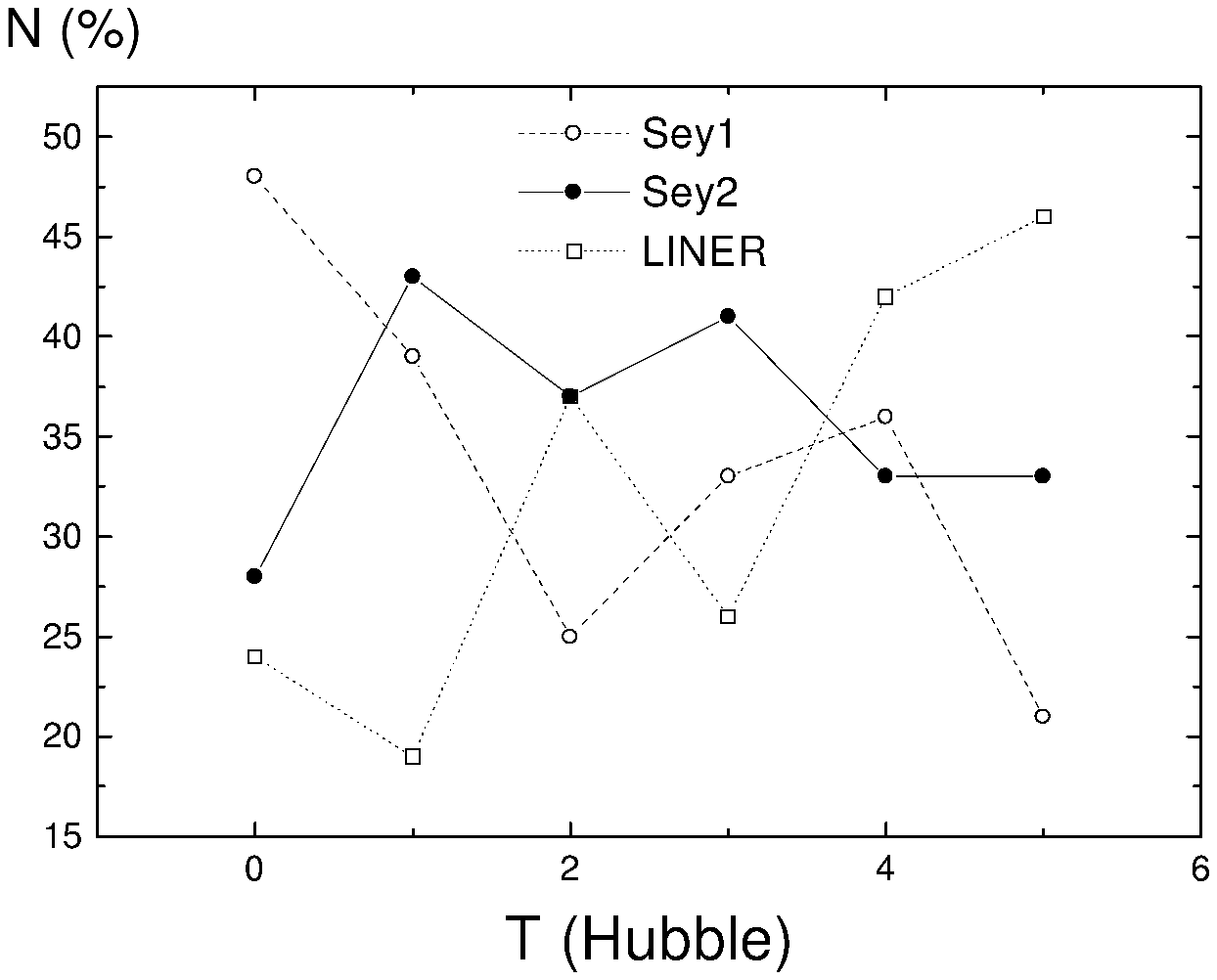}
\caption{Left: frequency of strongly barred galaxies (B's), weakly barred galaxies (X's) and all
barred galaxies (B's + X's) along the Hubble sequence of all galaxies brighter than 14 B mag and
with a bar classification in the RC3 (\cite{deVaucouleurs91}). Right: frequency of all Seyferts
(1 and 2) and LINERs along the Hubble sequence in the catalog of \cite{Lipovetsky88}.}
\end{figure}

In this regard, it is interesting that we have found that weak bars have bluer colors
than strong bars. This suggests that the former are in the process of formation rather than
in dissolution. To further analyse this issue, we have studied the frequency of bars along the Hubble
sequence, and found an excess of weakly barred galaxies in morphological classes T=5$\pm$1,
which corresponds to Sbc, Sc and Scd. Moreover, we have also found that, while
Seyferts are more common in early-type spirals (T=1), LINERs are preferentially found in late-type
ones (T=5, see Figure 2). If we distinguish, among galaxies with AGN, those which are interacting
(i.e., either in pairs, groups or clusters) and those which have bars, like in Table 1, we can see
that early-type galaxies are mostly interacting, and that the fraction of barred galaxies rises
when we consider late-type galaxies. Taken altogether, these results seem to indicate that
young bars appear more often and are more important to feed AGN (especially LINERs) in late-type
spirals, rather than in early-type ones, where interactions seem to play a more relevant role.

\begin{table}
\begin{center}
\begin{tabular}{ccc}
T & Interacting & Barred \\ \\
0 - 1 & 68 (56\%) & 35 (29\%) \\
2 - 3 & 51 (53\%) & 34 (35\%) \\
4 - 5 & 25 (44\%) & 26 (46\%) \\
\end{tabular}
\caption{Galaxies with AGN which are either interacting and/or have bars in the catalog of
\cite{Lipovetsky88}. The numbers in parentheses are the fraction of galaxies satisfying these conditions
in each morphological bin.}
\end{center}
\end{table}

\begin{acknowledgments}
DAG would like to express his gratitude to the organizers of this meeting.
This work was financially supported by FAPESP grant no. 99/07492-7.
\end{acknowledgments}

\end{document}